# Identifying Exceptional Points in Two-Dimensional Excitons Coupled to an Open Optical Cavity

*Ben Johns[§]\*, Nitin Yadav, Anand Vinod[#], Kuljeet Kaur, and Jino George\**

Molecular Strong Coupling Lab, Department of Chemical Sciences, Indian Institute of Science Education and Research (IISER) Mohali, Punjab, India - 140306.

[§]Current affiliation: Ultrafast Nanoscience Unit, Department of Physics, Umeå University, Linnaeus väg 24, 90187 Umeå, Sweden

[#]Current affiliation: CNR Nanotec, Institute of Nanotechnology, 73100 Lecce, Italy

**ABSTRACT: Strong coupling in the conventional sense requires that the Rabi cycling process between two interacting states is faster than other dissipation rates. Some recent experimental findings show intriguing properties that were attributed to polaritonic states (e.g., plexcitons) even though the above criterion is not satisfied. Here, we theoretically predict and provide experimental evidence of polariton-like behavior in a system that does not show Rabi splitting. The photoluminescence of an exciton-cavity system consisting of a two-dimensional exciton monolayer (tungsten disulfide, $WS_2$) coupled to a planar, open, one-mirror optical cavity configuration is studied. We experimentally observed a transition from the weak coupling regime crossing an exceptional point to form polariton-like states by varying the coupling strength and the cavity loss. Our observations are fully in agreement with a theoretical quasi-normal mode analysis, which predicts this transition and confirms the presence of exceptional points in the system. These results**

**provide evidence that polaritonic effects can be experimentally observed even when the conventional strong coupling condition is not satisfied.**

## INTRODUCTION

In strongly coupled systems, the hybrid states have half-light and half-matter characteristics, called polaritonic states.[1] Recently, these systems have received much attention because of their intriguing properties, such as condensation,[2] lasing,[3] cavity-modified conductivity,[4] energy transport[5] and chemical reactivity.[6] These interesting properties have led to discussions and differing opinions about the conditions to be satisfied for a system to reach strong coupling. Typically, the emergence of a Rabi splitting in the optical spectrum is recognized as a signature of strongly coupled interactions. To facilitate a clear distinction between the weak and strong coupling regimes, the condition that the rate of energy exchange is greater than the decoherence rates ($g > (\Gamma_C + \Gamma_X)/4$) is widely adopted.[7–10] Here, $g$ denotes the light-matter coupling strength, and $\Gamma_C/2$ and $\Gamma_X/2$ are the decoherence rates of the cavity and exciton, respectively. However, the condition for the existence of polaritonic states is different; this is given by $g > |\Gamma_C - \Gamma_X|/4$ (refs. [8,11]). This is related to the emergence of non-degenerate eigen energy levels in the coupled system.[12] Notably, recent works on polaritonic behavior in plexcitonic systems have adopted this relation to represent the strong coupling condition.[13,14]

In interacting light-matter systems, the weak and strong coupling regimes are separated by a so-called exceptional point (EPs).[8,15,16] An EP is a point in the parameter space of a physical system where its eigenmodes and corresponding eigenvalues coalesce and become indistinguishable[12]. A physical system operating in the vicinity of an EP can exhibit exotic properties such as enhanced optical sensing,[16,17] lasing,[18,19] topology-protected transport,[20] and all-optical modulation.[21] Typically, EPs in polaritonic systems are detected by the emergence of a spectral Rabi splitting as the system transitions from the weakly coupled regime to the strongly coupled regime. Such a transition can be brought about by tuning the energies[16], coupling strength[22] or intrinsic losses[12] of the interacting resonators. When the resonators are degenerate in energy, their optical absorption spectrum displays a transition from a single feature (peak or dip) to split features, which is then identified as Rabi splitting.[1,23] Note that this is different from a spectral splitting that can arise in the weakly coupled regime e.g., in the case of electromagnetically-induced transparency.[24,25] Recent results have shown that the observed Rabi splitting may not always represent the actual level splitting (LS).[26] In other words, the observed peaks or dips may not directly correspond to the eigenenergy values of the polaritonic states represented by the LS. There are many reports that directly use linewidth observations to access and manipulate the strength of strong coupling.[27] Notably, a non-zero LS can be hidden by the presence of losses in the system.[28] This is especially relevant in open polaritonic systems[4,21,29–32] where the coupling strength is not large enough compared to the losses, and a small but non-zero LS may be masked by the linewidths of the interacting systems. Therefore, the identification of exceptional points and polaritonic states in such classes of interacting systems requires a careful assessment of the LS going beyond a direct spectral read-out of the Rabi splitting.

Here, we experimentally show and theoretically validate the presence of polaritonic states in a system that cannot be inferred from Rabi splitting measurements. We explore a model system consisting of a two-dimensional semiconducting monolayer (tungsten disulfide, $WS_2$) coupled to a planar, open, one-mirror cavity formed by a thin layer of hexagonal boron nitride (hBN) on a mirror. We demonstrate that by suitably tuning the thickness of hBN, the weakly coupled exciton-cavity system transitions across an EP to enter the strong coupling regime. We also observed experimentally a transition in the emission from exciton-like to polariton-like behavior as the system crosses the EP, even though there is no measurable Rabi splitting in the absorption or emission spectra. This is further validated by theoretical quasi-normal mode analysis that extracts the LS and clarifies the presence of polaritonic states in this system. These results help us to identify EPs and polaritonic states and present a platform for fundamental studies into polaritonic phenomena in open optical systems.

## RESULTS AND DISCUSSION

The energy levels of an exciton $(E_X - i\Gamma_X/2)$ coupled to an optical cavity $(E_C - i\Gamma_C/2)$ are expressed in terms of a real part corresponding to the resonant energy $(E_X, E_C)$ and an imaginary part representing the damping (full width at half maximum) of the oscillations $(\Gamma_X, \Gamma_C)$. The eigenenergy levels of the coupled non-Hermitian system for a given detuning between the exciton and cavity $\delta = E_C - E_X$ can be written as[12,33]

$$E_\pm = \tilde{E}_{av} \pm \Omega_{LS}/2,$$

where $\tilde{E}_{av} = (E_C + E_X)/2 - i(\Gamma_C + \Gamma_X)/4$ is the average energy of the exciton and cavity states.[34] The detuning between the eigenstates, called level splitting (LS),[26,33] is given by $\Omega_{LS} = \sqrt{4g^2 + \left[\delta + i\frac{(\Gamma_C - \Gamma_X)}{2}\right]^2}$. Here, we use the term LS to contrast with Rabi splitting, which denotes the spectroscopically measured energy splitting. The exciton-cavity interaction is characterized by the coupling strength $g$, which represents the rate of energy exchange between light and matter states. The coupling strength depends upon the transition dipole moment of the exciton $\boldsymbol{\mu}$, the electric field of the cavity $\boldsymbol{E}$, and their relative orientations, i.e., $g = \boldsymbol{\mu}.\boldsymbol{E}$. Note that in the case where $N$ identical excitons are coupled to a cavity as seen under typical experimental conditions, the coupling strength scales as $\sqrt{N}g$. When $\delta = 0$, the cavity and exciton modes are resonant $(E_X = E_C \equiv E_0)$ and the level splitting becomes

$$\Omega_{LS} = 2\sqrt{g^2 - g_{EP}^2},$$

where $g_{EP} = |\Gamma_C - \Gamma_X|/4$ is the coupling strength at the EP.[33] Depending on the sign inside the square root, the interaction of the system can be classified into strongly coupled $(g > g_{EP})$ and weakly coupled $(g < g_{EP})$ regimes separated by an EP at $g = g_{EP}$. In the strong coupling regime where $\Omega_{LS}$ is real, the eigenenergies have distinct real parts and form polaritonic states with energies given by $\Re(E_\pm) = E_0 \pm \Omega_{LS}/2$ and $\Im(E_\pm) = -(\Gamma_C + \Gamma_X)/4$ (solid and dashed lines respectively in **Figure 1a**). On the other hand, in the weak coupling regime, $\Omega_{LS}$ is imaginary ($\Omega_{LS} = i\Delta_{LS}$, where $\Delta_{LS}$ is real), and the real eigen energies are identical with $\Re(E_\pm) = E_0$ while the imaginary parts become distinct, $\Im(E_\pm) = -(\Gamma_C + \Gamma_X)/4 \pm \Delta_{LS}/2$. At the EP, the two eigen energies merge and become indistinguishable, $E_\pm = E_0 - i(\Gamma_C + \Gamma_X)/4$, representing a singularity of the non-Hermitian system, as represented in **Figure 1a**.[12,34]

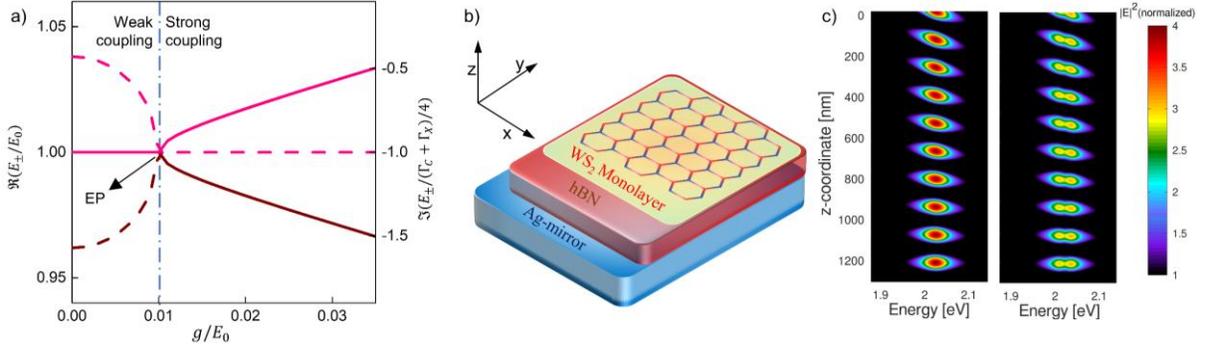

**Figure 1:** (a) Real part $\Re(E_\pm)$ normalized to $E_0$ (solid lines, left axis) and imaginary part $\Im(E_\pm)$ normalized to $(\Gamma_C + \Gamma_X)/4$ (dashed lines, right axis) of the eigenenergies of a representative coupled non-Hermitian system at $\delta = 0$ with $g_{EP} = 0.01 E_0$ and $\Gamma_C + \Gamma_X = 0.07 E_0$. The location of the EP where the eigen-energies merge is marked by an arrow. (b) Schematic of the investigated open cavity structure and (c) the field distribution for the empty cavity (left) and the exciton-cavity (right). In (c), $z = 0$ indicates the position of WS$_2$ and $z = 1250$ nm corresponds to the bottom of the hBN layer.

We explore the occurrence of such a weak-to-strong coupling transition across an EP in our lithography-free platform consisting of a monolayer of WS$_2$ separated from an opaque silver mirror by an hBN flake of thickness $L$ (**Figure 1b**). The excitonic transition dipoles in monolayer TMDs orient randomly in the 2D plane (*x-y*) owing to an ultrafast decoherence process[35] and have a negligible out-of-plane component[36] ($\boldsymbol{\mu} = \mu_{xy}$). To maximize the light-matter interaction in our open system, we probe the coupling of the A-exciton of WS$_2$ at 2.02 eV (refs. [35–39]) with the TE modes of the open cavity, which have electric fields only in the *x-y* plane ($\boldsymbol{E} = E_{xy}$). **Figure 1c** shows the calculated field distribution (using transfer matrix method, see **Materials and Methods**) within the dielectric medium in an hBN-on-silver cavity without WS$_2$ ("empty cavity", left panel) and with WS$_2$ ("exciton-cavity", right panel). Note that by placing the monolayer on top of hBN, we ensure that it experiences the maximum electric field of the empty cavity. This is also visualized by the splitting in the field distribution of the exciton-cavity. This property arises due to the open boundary condition of the empty cavity at the hBN-air interface whereby a field maximum occurs at the boundary. This is an attractive and extremely useful practical feature of our open system, which contrasts with conventional Fabry-Perot structures where the monolayer must be placed in the interior to maximize light-matter interactions.

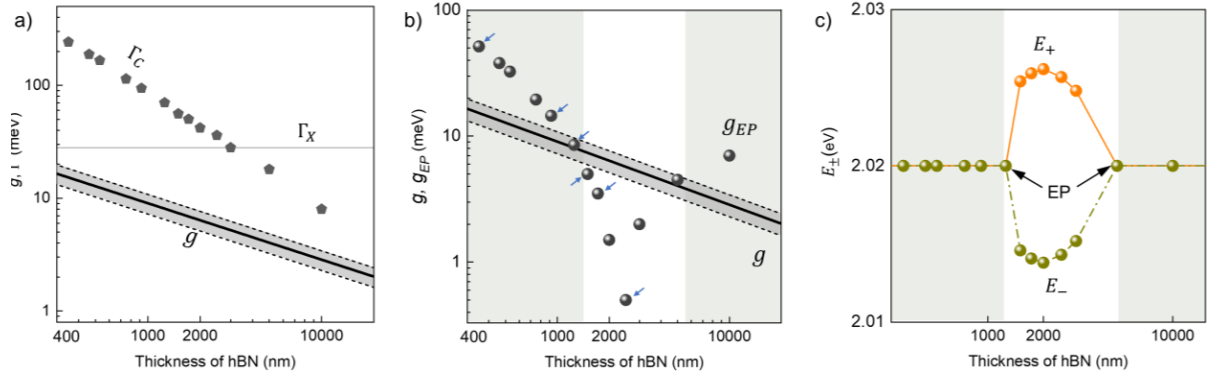

**Figure 2:** (a) Variation of $g$ and $\Gamma_C$ with hBN thickness, $L$ for the hBN-on-silver cavity. The value of $\Gamma_X$ is marked for comparison. The uncertainty in the calculated values of $g$ is also indicated. (b) Comparison of $g$ and $g_{EP}$ as a function of hBN thickness. The shaded and unshaded regions indicate thicknesses where $g < g_{EP}$ and $g > g_{EP}$, respectively. The samples studied experimentally are identified with arrows. (c) Real part of the eigenenergy values of the exciton-cavity system calculated from Equation 1. The presence of two EPs bounding a region of non-zero level splitting can be seen.

To predict the existence of the EP, we theoretically calculate the coupling strength $g$ and the critical coupling strength $g_{EP}$ of our system. **Figure 2a** plots the variation in $g$ and $\Gamma_C$ with $L$. The value of $\Gamma_X$ for the A-exciton in WS$_2$ is also shown for comparison. We note that both $g$ and $\Gamma_C$ varies inversely with $L$. Since the cavity linewidth is inversely proportional to the round-trip time ($t \sim 2L/c$, where $c$ is the speed of light), cavities with greater thickness possess a lower $\Gamma_C$. Moreover, the cavity mode field can be expressed in terms of the mode volume $V$ as $\boldsymbol{E} = \sqrt{\hbar\omega/2\epsilon_r\epsilon_0 V}$, where $\omega$ is angular mode frequency, $\epsilon_r$ is the relative permittivity of the cavity medium, and $\epsilon_0$ is the permittivity of free space. Given that the WS$_2$ layers are identically placed at the anti-node position for all thicknesses, the coupling strength scales inversely with the mode volume as $g \propto 1/\sqrt{V}$, and hence $g$ shows an inverse dependence on $L$. However, the overall decrease in $g$ with hBN thickness is much smaller compared to the change in $\Gamma_C$, as observed in **Figure 2a**. **Figure 2b** shows the corresponding comparison of $g$ with $g_{EP}$. Evidently, $g_{EP}$ initially decreases when $\Gamma_C > \Gamma_X$, reaches a minimum at $\Gamma_C = \Gamma_X$, and increases again when $\Gamma_C < \Gamma_X$. As a result, we see that $g > g_{EP}$ in the range approximately 1100 nm $\leq L \leq$ 5000 nm. Therefore, the system is predicted to possess two EPs (at $L \approx$ 1100 nm and 5000 nm) on either side of the condition $\Gamma_C = \Gamma_X$, and enters the strong coupling regime bounded by the two EPs (**Figure 2c**). **Figure 2** presents an interesting instance where polaritonic states can be present when $g < \Gamma_C, \Gamma_X$.[33] Consequently, in contrast to typical systems where the focus is on maximizing the coupling strength between emitter and cavity, we see that whether our open system can enter the strong coupling regime depends strongly on the relative losses of the emitter and the cavity. This is also because the cavity losses (ranging from 5 – 200 meV) are highly sensitive to changes in $L$ while the coupling strength (5 – 20 meV) does not vary much with thickness.

To experimentally probe the existence of EPs and polaritonic states in this system, we prepared samples with a range of different thicknesses across the first (low thickness) EP at $L$ = 1100 nm. The experimentally studied thicknesses are identified with arrows in **Figure 2b**. We probe the samples using Fourier plane micro-spectroscopy in reflection and PL emission modes (see **Materials and Methods** for sample preparation and measurement details). **Figure 3a-d** shows the measured angle-resolved spectra collected in reflection mode for four selected hBN thicknesses. The WS$_2$ monolayer absorbs strongly due to the A-exciton signature at 2.02 eV as indicated by the horizontal lines in **Figure 3**. As $L$ increases from 475 nm to 1720 nm, the monolayer interacts with the higher order modes of the cavity, from 4$^{th}$ mode in **Figure 3a** to 11$^{th}$ and 12$^{th}$ modes in **Figure 3d**. A strong cavity-mediated absorption effect is found in all the samples at the exciton energy as we observed previously[38], featuring a pronounced angle-selective reflectance dip where the cavity mode and the exciton are degenerate. However, no signature of strong coupling effects such as Rabi splitting is observed for any thickness, indicating that the samples may be in the weak coupling regime. The absence of splitting in reflection is verified by transfer matrix method (TMM) simulations for the three cavities. **Figure 3e-h** shows the corresponding spectra collected in PL mode. For $L$ = 475 nm, a flat PL signal is obtained at the emission wavelength of bare $WS_2$. A slight spectral broadening is observed compared to that of the bare monolayer emission, which is attributed to the presence of the broad cavity mode. In our earlier work, we showed that the cavities with lower $L$ also show a flat emission profile.[38] However, the remaining samples reveal markedly different features. For $L$ = 1055 nm, the PL signal displays an angle dependence that resembles the cavity dispersion observed in reflection mode. The effect of the cavity is also evident in the Purcell-enhancement at oblique angles ($k_\parallel \approx \pm 5\ \mu m^{-1}$) where the cavity and exciton become degenerate. However, the spread in the energy indicates that the emission is not purely excitonic in nature, but is not quite polaritonic emission either. It is interesting to note that this sample lies very close to the EP. We note that earlier works have suggested that such emission showing a mixed behavior may occur in systems in an intermediate coupling regime[40]. A similar behavior is observed for $L$ = 1355 nm with a more pronounced dispersive emission, which lies near the EP in the strong coupling side. Finally, the PL signal shown in **Figure 3h** for $L$ = 1720 nm reveals sharp polariton-like dispersive emission characteristics. A Purcell-enhanced emission is observed at the exciton energy which is accompanied by two new branches at higher and lower energy. In comparison to the previous samples, the emission here is narrower and closely follows the dispersion of the cavity modes observed in reflection. The PL signal plotted on a logarithmic scale further confirms the presence of three dispersing branches in the emission. The broadband, dispersive emission of this sample does not appear to be $WS_2$ emission simply filtered through the cavity. To support this, comparison with the emission of an isolated $WS_2$ monolayer shows that the overall emission from the exciton-cavity is distributed over a much larger energy range than the energy spread of emission from $WS_2$. Thus, we infer that the PL signal arises from polariton-like branches formed by two cavity modes and the $WS_2$ exciton. To the best of our knowledge, no previous work has reported the observation of a dispersive polaritonic emission signal in the absence of Rabi splitting in the system. Moreover, polaritonic emission from a monolayer 2D exciton coupled to an unpatterned open optical cavity has not been observed previously. Even though no Rabi splitting is observed in reflection (as verified by TMM simulations) or emission mode, the PL signals from the exciton-cavities reveal a transition from bare $WS_2$-like emission to polaritonic emission with varying hBN thickness.

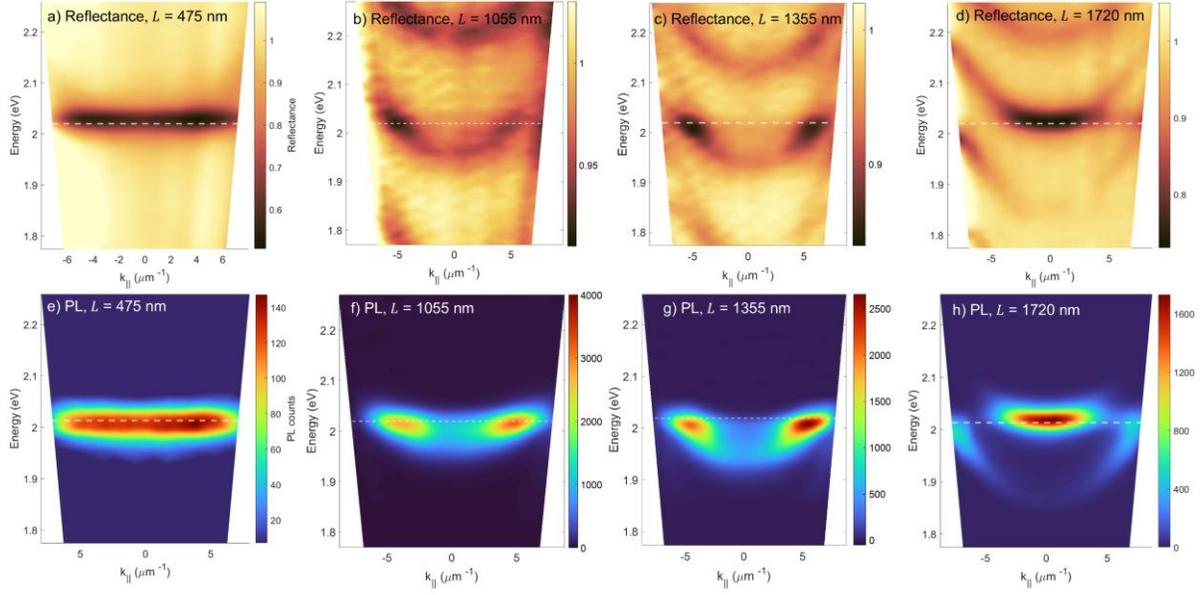

**Figure 3**: (a-d) Reflection mode and (e-h) PL emission mode dispersion for four exciton-cavity samples of thicknesses $L$ = 475 nm, 1055 nm, 1355 nm, and 1720 nm.

Although the weak-to-strong coupling transition observed here was predicted by the analysis in **Figure 2**, the absence of Rabi splitting casts doubts on the nature of strong coupling in our system. To further elucidate the underlying light-matter interactions and understand our experimental observations, we calculate the LS of the coupled cavities using a quasi-normal mode (QNM) analysis.[41,42] Quasi-normal modes are the eigenmodes of non-Hermitian systems and are intrinsic properties useful in describing optical systems with dissipation or radiation losses. Assuming an $\exp(-i\omega t)$ convention, the complex angular frequency $\widetilde{\omega}_m$ of

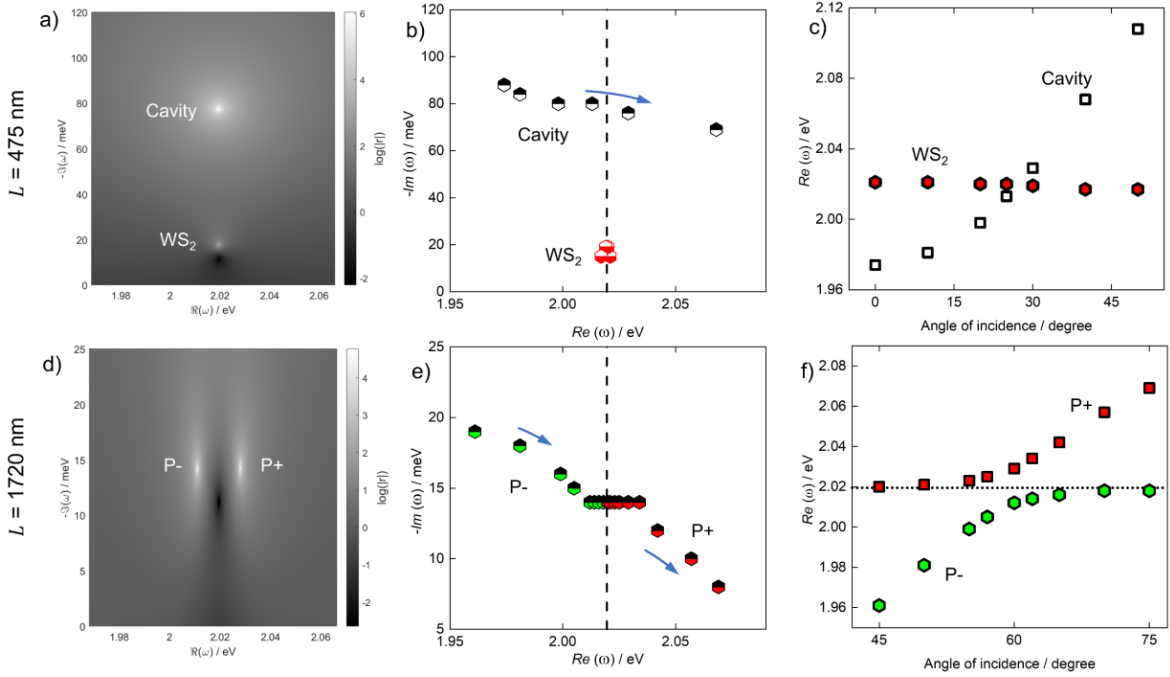

**Figure 4**: (a-c) Weakly coupled sample $L$ = 475 nm. (a) Map of log($|r|$) when the cavity mode and WS$_2$ exciton are degenerate (b) Trajectory of modes in the complex frequency plane. The arrow indicates the direction of increasing angle of incidence. The energy of the WS$_2$ A-exciton is indicated by the vertical line. (c) Variation in the real part of QNM eigenfrequencies $\Omega_m$ with angle of incidence showing the crossing of modes. (d-f) Strongly coupled sample $L$ = 1720 nm. (d) Map of log($|r|$) when the cavity mode and WS$_2$ exciton are degenerate showing a non-zero level splitting. (e) Trajectory of modes in complex frequency plane showing evolution of P- and P+ states, (f) Dispersion of $\Omega_m$ showing anti-crossing behavior at the exciton energy shown by the horizontal line.

a QNM is given by $\hbar\widetilde{\omega}_m = \Omega_m - i\Gamma_m/2$, where $\Omega_m$ is the resonance energy of the mode, $\Gamma_m$ is the energy loss rate of the mode (full-width at half-maximum), and $m$ denotes the mode order.[42] The $\hbar$ has been added to indicate that we use $\Omega_m$ and $\Gamma_m$ in units of energy (eV). Numerically, we identify the QNM eigenfrequencies of our system from the location of the poles of its reflection coefficient, $r$ (ref.[43]). We adopt a graphical pole-search method in the complex-frequency plane $(\Re(\omega), \Im(\omega))$ using TMM to identify $\widetilde{\omega}_m$. **Figure 4a** plots log($|r|$) in the complex-frequency plane for the weakly coupled exciton-cavity with $L$ = 475 nm, at an angle of 27 degrees. This corresponds to $k_{\parallel}$ = 4.6 $\mu m^{-1}$ in **Figure 3a** where the cavity and exciton are degenerate. Two poles (bright points) corresponding to the location of two degenerate QNMs are observed in the log($|r|$) plot, at a value of 2.02 eV on the $\Re(\omega)$ axis. The QNMs associated with the cavity and WS$_2$ are easily distinguishable here by comparing their locations on the $\Im(\omega)$ axis, i.e., their losses. For the lower QNM, $\Im(\hbar\widetilde{\omega}_m) = 14$ meV, which is readily identified as the half-width of the A-exciton in a WS$_2$ monolayer.[44] The QNM with $\Im(\hbar\widetilde{\omega}_m) \approx 80$ meV is then identified as the cavity mode. This agrees with the estimate of $\Gamma_C$ for the corresponding empty cavity in **Figure 2a**. **Figure 4b** plots the trajectory of the QNMs of the coupled system in the complex-frequency plane calculated by varying the angle of incidence. As expected under weak coupling, the QNM corresponding to the exciton remains unperturbed, while the cavity QNM crosses the exciton energy due to its dispersing nature. Please note that the cavity gets perturbed while crossing the absorption regime in the weakly coupled system. This is under the assumption that FHWM is affected as the cavity mode crosses the optical transition of the exciton. This weakly coupled nature in the $L = 475$ nm exciton-cavity is highlighted in **Figure 4c**, where the real parts of the QNM eigenfrequencies ($\Omega_m$) are plotted as a dispersion curve, showing that the mode energies cross each other without splitting. **Figure 4d-f** presents the corresponding plots for the $L = 1720$ nm exciton-cavity, which we theoretically predicted to be strongly coupled and experimentally exhibited polariton-like emission. The map of log($|r|$) in **Figure 4d** when the empty cavity and exciton are degenerate reveals the presence of two QNMs on either side of the exciton energy, with an energy separation of $\Omega_{LS} = 17$ meV. **Figure 4e** shows the trajectory of the modes in the complex-frequency plane by varying the angle of incidence. In contrast to the weak coupling case, it is not possible here to attribute the modes to the cavity or exciton separately. Indeed, the dispersion of the calculated eigenenergy values exhibit an anti-crossing at the exciton energy (**Figure 4f**). This allows us to identify the QNMs of the coupled system for $L = 1720$ nm as P+ and P- polaritonic states with LS of 17 meV on either side of the exciton energy, as labelled in **Figure 4d-f**. Finally, we note that EPs occur when the two poles merge ($\Omega_{LS} = 0$ and $\Delta_{LS} = 0$).[45] For the $L$

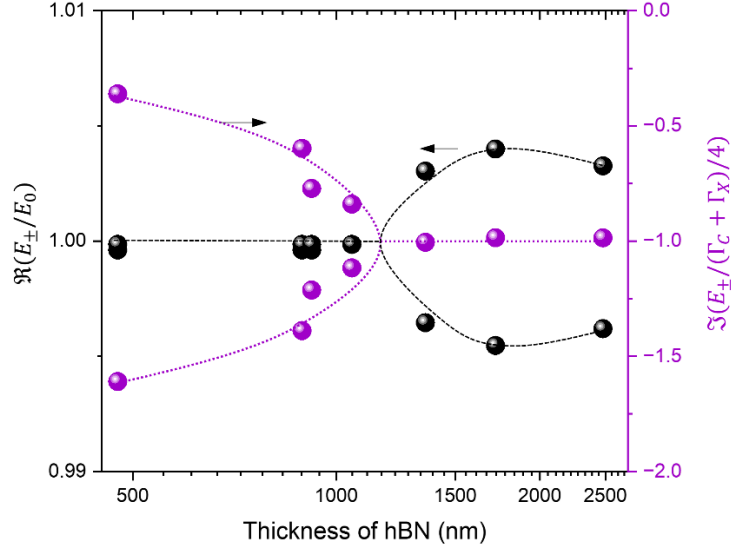

**Figure 5:** Real (left axis) and imaginary (right axis) parts of eigenenergies of the exciton-cavity obtained from QNM analysis. The same normalization for the eigenenergies values used in **Figure 1** is applied here. The dotted lines are a guide to the eye. $E_0$ is the A-exciton energy (2.02 eV).

= 1055 nm cavity (intermediate case), we see that $\Omega_{LS} = 0$ and $\Delta_{LS} \sim 7$ meV, i.e., the sample lies in the weak coupling regime ($g < g_{EP}$) but in very close proximity to the EP around $L = 1100$ nm as expected. On the other hand, for the $L = 1355$ nm cavity, $\Omega_{LS} \sim 14$ meV and $\Delta_{LS} = 0$ meV, i.e., the sample lies in the strong coupling regime ($g > g_{EP}$) near the EP.

**Figure 5** reports the results of QNM analysis on all the experimentally investigated samples. The real and imaginary parts of the normalized eigenmode energies are plotted, evidencing the weak to strong coupling transition around $L = 1100$ nm thickness of hBN. When $L < 1100$ nm, $\Omega_{LS} = 0$ i.e. weak coupling, while when $L > 1100$ nm, $\Omega_{LS} > 0$ i.e. strong coupling. Further, we remark that the QNM analysis is in excellent agreement with our predictions in **Figure 2** made using simple arguments. Our analysis demonstrates the presence of polaritonic states for the strongly coupled exciton-cavity ($L = 1720$ nm) and clarifies the polaritonic origin of light emission observed in our experiments. It also indicates that intermediate coupling behavior in this system is the result of its proximity to the EP. An important question remains: why is there no visible Rabi splitting observed experimentally when both QNM analysis and emission mode data indicate the presence of polariton-like states? To understand this, we first note that in the strongly coupled exciton-cavity ($L = 1720$ nm), $g_{EP} \approx 0$, making it extremely efficient for the system to enter the strong coupling regime ($g > g_{EP}$). However, we find that $\Omega_{LS} < (\Gamma_X + \Gamma_C)/2$, i.e., strong coupling in the conventional sense where the rate of energy exchange is greater than the dissipation rates is not present here. In other words, the LS is lower than the linewidths of the interacting systems as noted earlier, making the splitting too weak to be observable by optical spectroscopy.[28] This regime has been referred to as intermediate coupling, or pseudo-strong coupling by various authors.[26,28,46] Earlier works have also pointed out that the splitting observed in reflection or transmission spectra may not exactly correspond to the polariton energies.[47] However, we stress that systems previously identified to be in intermediate coupling regimes have not been observed to exhibit polaritonic behavior such as the dispersive emission we observe here.[48] On the other hand, there are multiple observations that report a splitting in reflection or transmission spectra but do not see polaritonic emission, e.g., plexcitonic systems.[49] This has been suggested to occur in systems that have not fully entered the strong coupling regime.[40] As a result, the observation of a splitting in emission is considered to be stronger evidence for polariton formation than the observation of a splitting in reflectance.[49,50] However, to the best of our knowledge, no system has been previously reported that displays polaritonic (dispersive) emission in the absence of Rabi splitting. This points to a potentially significant and unique behavior of this system: the coupling strengths required to observe polaritonic effects such as dispersive emission is extremely low here compared to other exciton-polariton systems reported previously. This opens avenues for utilizing open cavity polaritonic effects in novel applications, including polaritonic chemistry, lasing, quantum sensing, etc. Further, our results also shed new light on the properties of open polaritonic systems. For instance, it supports the possibility that polaritonic effects may play a much larger role than previously thought in open cavity-coupled systems and in processes such as photosynthetic energy transfer in natural systems or light emission in solid state LEDs.[30] Recent works have already demonstrated such effects in polaritonic devices, including improved conductivity in open systems.[31]

Finally, to explore whether measurable Rabi splitting can be achieved with our open cavities, we consider the possibility of realizing $g > \Gamma_C, \Gamma_X$, which is the case typically in conventional strong coupled systems.

For this, we place a thick WS$_2$ multilayer (~ 200 layers) on an $L$ = 3300 nm cavity. As $g \propto \sqrt{n}$ (where $n$ is the number of layers) and assuming that the dipole moment of an exciton in the thick multilayer flake is half that in a monolayer,[36] we find that the coupling strength for $N$ = 200 is ≈ 7 times greater than that for a monolayer. This gives a rough estimate for $g$ as 35 meV, and the estimated Rabi splitting is $\Omega_R \approx 2g$ = 70 meV. The measured reflectance of this system displays a pronounced Rabi splitting of 60 meV, which is close to the theoretical estimate above. Please note that the emission features disappear for multilayer 2D materials and therefore cannot be probed. The observed splitting is also captured by the TMM simulations, verifying that the conventional regime of strong coupling with a measurable Rabi splitting can also be achieved in our open cavity system using a multilayer 2D material.

**CONCLUSIONS**

The conventional definition of strong coupling is Rabi cycling overtaking any other dissipation rates in the system and, therefore, the observation of polaritonic branches. However, this definition doesn't hold well in many systems, such as plexcitons and open cavities. This work presents the observation of polaritonic effects in 2D excitons coupled to an open optical cavity. Even though the cavity is leaky, it is possible to reach the strong coupling condition by varying the cavity loss and coupling strength via cavity thickness tuning. Experimental measurements do not reveal measurable Rabi splitting; however, polariton-like emission is observed from samples that allowed us to explore other methods like level splitting model to investigate the origin of this behavior. Please note that emission measurements are more sensitive probes in such conditions. We were also able to identify samples in the vicinity of an exceptional point from its emission profile that exhibits intermediate behavior between that of the exciton-like and polariton-like states. Analysis of the coupled samples in terms of QNMs supports the experimental observations and identifies the presence of a weak-to-strong coupling transition across the exceptional point. Our observation of polariton-like states and exceptional points in a simple open cavity system strengthens the possibility of exploiting the properties of 2D exciton-polaritons in lithography-free platforms that can be used in integrated photonic systems. Our work also highlights the usefulness of QNM analysis in elucidating the underlying light-matter interaction regimes in polaritonic systems.[51] We propose that our system may be an ideal platform for ultrafast dynamics studies to shed more light onto the fundamental behavior of polaritons, especially in the vicinity of exceptional points.

**MATERIALS AND METHODS**

**Sample preparation**

The sample preparation details are given in detail in ref [[38]]. Briefly, an opaque silver mirror (120 nm) is sputtered on a glass substrate. An hBN layer is then mechanically exfoliated from bulk hBN onto a PDMS polymer substrate using scotch tape and its thickness is estimated from the interference fringes in its transmission spectrum. Finally, the hBN layer is transferred onto the silver mirror using a custom-built transfer stage. Freshly exfoliated $WS_2$ monolayers are subsequently transferred onto the hBN on mirror structure using the same mechanical exfoliation and transfer process.

**Reflection mode and emission mode in Fourier plane micro-spectroscopy**

Reflection mode data were obtained using a halogen lamp with a broadband spectrum covering the visible range. In emission mode, the 434 nm line of a mercury lamp is selected using a 10 nm bandpass filter and used as the excitation source. The broadband or narrowband sources are focused onto the sample using a 100X, 0.75 N.A. objective in a Nikon inverted microscope (Eclipse Ti2) at room temperature, which also collects back the reflected or emitted light. A 4F relay optics system using a Fourier lens was used to measure angle-dependent reflection and emission of the exciton-cavity samples. Two achromatic lenses (f = 100 mm) relay the image focused by the microscope objective and tube lens onto the spectrometer CCD array (SpectraPro HRS-300, Princeton Instruments and Pylon 100BX, cooled to −120 °C using liquid nitrogen for spectral acquisition). An additional f = 50 mm achromatic lens (Fourier lens) was introduced into the optical path before the CCD slit to image the Fourier plane. The 0.75 N.A. objective allows the collection of angular range up to 50°, which was resolved in the Fourier plane.

**Transfer matrix method calculations**

The electric field distribution, reflectance, and QNM analysis of the planar structure is calculated using a custom-written transfer matrix method code outlined in detail in the supporting information of ref. [52]

**AUTHOR INFORMATION**

**Corresponding Authors**

Jino George – Indian Institute of Science Education and Research (IISER), Mohali, Punjab 140306, India; Email: jgeorge@iisermohali.ac.in; orcid.org/0000-0002-3558-6553

Ben Johns – Indian Institute of Science Education and Research (IISER), Mohali, Punjab 140306, India (Current affiliation: Ultrafast Nanoscience Unit, Department of Physics, Umeå University, Linnaeus väg 24, 90187 Umeå, Sweden)

Email: ben.johns@umu.se;  ben.johnsj@gmail.com  orcid.org/0000-0002-6752-5931

**Authors**

Nitin Yadav – Indian Institute of Science Education and Research (IISER), Mohali, Punjab 140306, India.

Anand Vinod – Indian Institute of Science Education and Research (IISER), Mohali, Punjab 140306, India. (Current affiliation: CNR Nanotec, Institute of Nanotechnology, 73100 Lecce, Italy)

orcid.org/0009-0003-0877-297X

Kuljeet Kaur – Indian Institute of Science Education and Research (IISER), Mohali, Punjab 140306, India; orcid.org/0009-0003-6190-5604.

**Author Contributions**

The manuscript was written through the contributions of all authors. All authors have given approval to the final version of the manuscript.

**Funding Sources**

The authors thank the SERB-core research grant (**CRG/2023/001122**), and IISER Mohali for the institute start-up grant.

**Notes**

The authors declare no competing financial interest.

ACKNOWLEDGMENT

B.J. thanks IISER Mohali for a Research Associate fellowship. N.Y., A.V. and K.K. thank IISER Mohali for their research fellowships.